\documentclass{article}

\usepackage{thanawat}
\usepackage{natbib, fullpage}

\title{Ex-Post Clock Auction}

\author{%
  Thanawat Sornwanee\\
  Stanford GSB
}

\begin{document}

\maketitle

\begin{abstract}
    We introduce an ex-post clock auction where the auctioneer gets to observe the value (or tail quantile in case of atomic distribution) of all but the highest bidder and use such information to price in an ex-postly optimal manner. 
    The ex-post clock auction can guarantee approximately 0.682156 fraction of the revenue of the optimal auction, and it can have revenue to be unboundedly multiplicatively bigger than that of the vanilla ascending clock auction.
\end{abstract}

For a distribution $F \in \Delta([0, \infty))$, we can define the inverse of survival function
\begin{align}
    S_F^{-1}(u)
    :=
    \sup\oo{\left\{
        p \in [0, \infty): F(p) \le 1-u.
    \right\}}
\end{align}
for any $u \in (0,1]$, and the revenue function, which is the posted price revenue as a function of tail quantile,
\begin{align}
\label{eqn:R}
    R_F(u) := 
        u S_F^{-1}(u)
\end{align}
for any $u \in (0,1]$.

In a conventional ascending clock auction with $n$ bidders with symmetric independent private value, the tail quantile is $U_i \overset{\text{iid}}{\sim} \text{Unif}([0,1])$, we observe the tail quantile $U_{(2:n)}, U_{(3:n)}, \dots, U_{(n:n)}$ where $U_{(i:j)}$ is the $i^\text{th}$ lowest value among $U_1, U_2, \dots U_j$. The auctioneer places a price equivalent to the tail quantile of $U_{(2:n)}$ in such ascending clock auction. However, we can see that the auctioneer, given such information, can profitably deviate to choosing the price corresponding to the tail quantile of
\begin{align*}
    U' \in \argmax_{u \in \oc{0, U_{(2:n)}}} \frac{1}{U_{(2:n)}}R_F(u)
\end{align*}
instead. We will see that this will still satisfy the ex-post incentive compatibility of the bidders similar to the conventional ascending clock auction or second price auction, but will further enjoys ex-post incentive compatibility for the auctioneer (under regularity).\footnote{This is done by adding the sequential agency back to the auctioneer similar to the framework in~\cite{sornwanee2026very}.} We denote this auction as ``ex-post clock auction" or ``symmetric ex-post clock auction".\footnote{We can see that in continuous distribution case, this will be equivalent to running a vanilla ascending clock to observe the valuation of every bidder except for the highest bidder. The auctioneer then gets to price such information.  However, if $F$ has some atom, a vanilla ascending clock will give rise to the event where everyone drops from the auction at the same time at the value of the atom $a \in \mathbb{R}^+_0$ with non-zero probability. If we use arbitrary tie-breaking, then ex-post incentive rationality will force the posted price to the winner to be at the price $P=a$. If the monopoly price for $F \vert_{> a}$ is strictly greater than $a$, this means that a bidder with value $V_i$ greater than but sufficiently close to $a$ will then want to deviate to report the value as $a$. Therefore, a more careful tie-breaking, and pricing is needed in order to ensure the strategyproofness of the auction. For example, the lookahead auction~\cite{ronen2001approximating} is a specific case of this where the tie-breaking is done arbitrarily to get a single winner among bidders of highest bids, and the posterior uses the identities and bids of all but the winner. However, since it is not a fully Bayesian framework, it remains unclear whether the additional information of tie-breaking can be utilized by the auctioneer. However, it is not clear which information is available to the auctioneer at the time of pricing.}

This suggests that we can increase the expected revenue from $\expect{\frac{R_F(U_{(2:n)})}{U_{(2:n)}}}$ to $\expect{\frac{\max_{u \in \oc{0, U_{(2:n)}}}R_F(u)}{U_{(2:n)}}}$. We will show that this increase can be infinitely large in multiplicative sense, and the ex-post clock auction, unlike ascending clock auction, is competitive with an optimal auction.

\section{Revenue Function}

    We first note that not every function $r:(0,1] \to [0, \infty)$ can be a revenue function $R_F$ for some distribution $F$. 

    \begin{corollary}
        [Characterization of Revenue Function]
        A function $R: (0,1] \to \mathbb{R}^+_0$ is a revenue function (according to the equation~\ref{eqn:R}) $R_F$ for some distribution $F \in \Delta\oo{[0, \infty)}$ if and only if all of the following holds:
        \begin{enumerate}
            \item $R$ is left-continuous;
            \item $\frac{R(u)}{u}$ is non-increasing in $u \in (0,1]$.
        \end{enumerate}
        We denote the collection of every function $R$ satisfying the conditions above as $\mathcal{R}_0$.
        
        Moreover, the mapping from $F \in \Delta([0,\infty))$ to $R_F \in \mathcal{R}_0$ is bijective.
    \end{corollary}
    \begin{proof}
        The only if part follows from that $S^{-1}$ is left-continuous by definition, and $\frac{R(u)}{u} = S^{-1}(u)$, which is non-increasing in $u >0$.
        The if part follows from that we can define $S^{-1}(u) = \frac{R(u)}{u}$ for $u\in (0,1]$, and finds a unique distribution $F$ with such $S_F^{-1} = S^{-1}$.
    \end{proof}
    As a simple result of this, we have the following corollary.
    \begin{corollary}
        For any $F \in \Delta([0, \infty))$, the revenue $R_F:(0,1] \to [0, \infty)$ is upper semi-continuous.
    \end{corollary}
    \begin{proof}
        This follows from that $\frac{R_F(u)}{u}$ is non-increasing and that $R_F$ is left-continuous.
    \end{proof}

    We can further require the regularity on $R_F$ so that an ex-post clock auction (ECA) exists or that an optimal clock auction (OPT) exists.
    \begin{proposition}
        [Necessary \& Sufficient Condition for the Existence of ECA]
        For any number of bidders $n \in \mathbb{N} -\{1\}$ and any distribution $F \in \Delta([0, \infty))$, a symmetric ex-post clock auction exists if and only if there exists some $\epsilon > 0$ such that $R_F(u)$ is non-decreasing in $u \in (0, \epsilon]$.
        For $n=1$, an ex-post clock auction exists if and only if $\argmax_{u \in (0,1]} R(u) \ne \emptyset$.
    \end{proposition}
    \begin{proof}
        If $n \ge 2$, then $U_{(2)}$ has full support, so the existence require $\argmax_{u \in (0,x]} R(u) \ne \emptyset$ for any $x \in (0,1]$. Since $R_F$ is upper semi-continuous, we then have that the necessary and sufficient condition is that $R_F$ is non-decreasing in the local neighborhood of $0$.
        
        For $n=1$, this is equivalent to the existence of a monopoly price.
    \end{proof}
    Note that this also requires that $R_F$ is bounded.

    The condition for the existence of the optimal auction is more involving.
    \begin{proposition}
        [Necessary \& Sufficient Condition for the Existence of OPT]
        For any number of bidders $n \in \mathbb{N} -\{1\}$ and any distribution $F \in \Delta([0, \infty))$, an optimal auction exists if and only if $\lim_{u \to 0^+} R_F(u) < \infty$ and, for any $q \in (0, 1]$, there exists some $q' \in (0,q)$ such that
        \begin{align*}
            \frac{R(q) - \lim_{x \to 0} R(x)}{q} \le \frac{R(q') - \lim_{x \to 0} R(x)}{q'},
        \end{align*}
        and $\argmax_{u \in (0,1]} R(u) \ne \emptyset$.
        
        For $n=1$, a symmetric ex-post clock auction exists if and only if $\argmax_{u \in (0,1]} R(u) \ne \emptyset$
    \end{proposition}
    \begin{proof}
        This follows from that the pooling region in optimal auction~\cite{myerson1981optimal}
    \end{proof}

    We define $\mathcal{F}$ to be the collection of value distribution $F \in \Delta([0,\infty))$ where optimal auction exists for every $n \in \mathbb{N}$. Thus, for any $F \in \mathcal{F}$, both optimal auction and ex-post clock auction exist.

\section{Ironings}

    For any $F \in \mathcal{F}$, $n \in \mathbb{N}$, we define
    \begin{align*}
        \text{OPT}(F,n)
    \end{align*}
    to be the expected revenue for the optimal auction with $n \in \mathbb{N}$ bidders under symmetric independent private value distributed according to the distribution $F \in \mathcal{F}$,\footnote{However, we can consider a bigger class of distribution to the one where $R$ is bounded and define $\text{OPT}(F,n)$ to be a supremum of expected revenue under any auction in the same setting satisfying Bayesian incentive compatibility and individual rationality.}
    \begin{align*}
        \text{ECG}(F,n)
    \end{align*}
    to be the expected revenue for the symmetric ex-post clock auction under the same setting,\footnote{Similarly, we can also define it as a limit case where the auctioneer choose a sequence of price that does not have to be an aposteriori monopoly price.} and
    \begin{align*}
        \text{SPA}(F,n)
    \end{align*}
    to be the expected revenue for a second price auction without reserve with $n \in \mathbb{N}$ bidders under the same setting.

    We can see that we can rewrite the revenue via two ironings.

    \subsection{Ironing via Monotone Majorant}

        Define $\bar{R}$ to be the least non-decreasing function $f: (0,1] \to [0, \infty]$ such that $f \ge R$ everywhere. This is well-defined, and we can do it in a position wise manner
        \begin{align*}
            \bar{R}(u) = \sup_{u' \in (0,u]} R(u')
        \end{align*}
        for all $u' \in (0,1]$.
    
    \subsection{Ironing via Monotone Concave Majorant}

        Define $\hat{R}$ to be the least concave function $f: (0,1] \to [0, \infty]$ such that $f \ge R$ everywhere. We have that the two operators, concave majorant and monotone majorant, are interchangeable. We further have that
        $\bar{\hat{R}}$ to be the least concave non-decreasing function $f: (0,1] \to [0, \infty]$ such that $f \ge R$ everywhere.

    The expected revenue for different auction under $F \in \mathcal{F}$ can then be simpy rewritten in the form of $R_F, \bar{R}_F,$ and $\bar{\hat{R}}_F$.

    We define an infinite collection of independent random variable
    \begin{align*}
        U_i \overset{\text{iid}}{\sim} \text{Unif}([0,1]),
    \end{align*}
    and we define ordered statistics for each $i, j \in \mathbb{N}$,
    \begin{align*}
        U_{(i:j)} := \inf\oo{\{u \in [0,1]: \sum_{k=1}^j \mathbf{1}_{U_k \le u} \ge i\}\cup \{1\}}.
    \end{align*}

    Therefore, we have that, for any $F \in \mathcal{F}$, $n \in \mathbb{N}$,
        \begin{align}
            \text{OPT}(F,n) &= \expect{\frac{\hat{\bar{R}}\oo{U_{(2:n)}}}{U_{(2:n)}}} = 
            \frac{1}{n}\expect{\hat{\bar{R}}\oo{U_{(1:n-1)}}},
        \end{align}
        \begin{align}
            \text{ECA}(F,n) &= \expect{\frac{\bar{R}\oo{U_{(2:n)}}}{U_{(2:n)}}}
            = 
            \frac{1}{n}\expect{\bar{R}\oo{U_{(1:n-1)}}}
            ,
        \end{align}
        and
        \begin{align}
            \text{SPA}(F,n) &= \expect{\frac{R\oo{U_{(2:n)}}}{U_{(2:n)}}}
            = 
            \frac{1}{n}\expect{R\oo{U_{(1:n-1)}}}.
        \end{align}

    A simple corollary is that
    \begin{corollary}
        $\text{OPT} \ge \text{ECA} \ge \text{SPA}$ everywhere.
    \end{corollary}
    \begin{proof}
        This follows from that $\hat{\bar{R}}_F \ge \bar{R}_F \ge R_F$ everywhere and for every $F \in \mathcal{F}$.
    \end{proof}

    Moreover, we have that an equality case exists. This can be seen from the case where $F = \delta_1$.

\section{Competitive Ratio}

    Next, we will show that we can get a bound on the competitive ratio between OPT and ECA.

    We define, for each $n \in \mathbb{N}$,
    \begin{align}
        \alpha_n := \sup_{t\in [0,1)}
        \frac{1+(n-1)t}{nt+(1-t)^n}.
    \end{align}

    We get the following property.
    \begin{lemma}
        $\alpha_n$ is non-decreasing in $n$, and 
        \begin{align}
            \lim_{n\to\infty}\alpha_n = \frac{1}{1-e^{-c}}
        \end{align}
        where $c$ is the unique positive solution of $e^c= c+2$.
    \end{lemma}
    \begin{proof}
        For each $n$, the mapping from $t \in [0,1) \mapsto \frac{nt}{1-t} \in [0, \infty)$ is bijective. Thus, by setting $x = \frac{nt}{1-t}$, we have that
        \begin{align*}
            \alpha_n = \sup_{x \in [0, \infty)} \frac{1+x}{x+\oo{1+ \frac{x}{n}}^{-(n-1)}}.
        \end{align*}

        Note that $(n-1)\log\oo{1+\frac{x}{n}} = \int_{s=0}^x \frac{n-1}{n+s} ds$, so $\oo{1+ \frac{x}{n}}^{n-1}$ is non-decreasing in $n$ and converges to $e^x$. Thus, for each $x \in [0, \infty)$, we have that 
        \begin{align*}
            \frac{1+x}{x+\oo{1+ \frac{x}{n}}^{-(n-1)}} \uparrow \frac{1+x}{x+e^{-x}}.
        \end{align*}
        Thus, $\alpha_n$ is non-decreasing.

        We also have that $\lim_{n\to\infty} \alpha_n = \sup_{x \in [0, \infty)} \frac{1+x}{x+e^{-x}}$. We can then get that it is $\frac{1}{1-e^{-c}}$.
    \end{proof}

    Moreover, we can define sequence $(t^*_n)_{n =2}^\infty$ such that
    \begin{align*}
        \left\{t^*_n\right\} := \argmax_{t \in [0, 1)} \frac{1+(n-1)t}{nt+(1-t)^n}.
    \end{align*}
    Note that this is well-defined since the objective function $\frac{1+(n-1)t}{nt+(1-t)^n}$ is continuous, is strictly quasi-concave, and has its limit to $1$, which is the same as when $t=0$. 

    Next, we will have the following theorem on the competitive ratio.
    \begin{theorem}
        For each $n \in \mathbb{N}$, we have that
        \begin{align*}
            \inf_{F \in \mathcal{F} -\{\delta_0\}} \frac{\text{ECA}(F,n)}{\text{OPT}(F,n)} = \frac{1}{\alpha_n},
        \end{align*}
        and
        \begin{align*}
            \inf_{F \in \mathcal{F} -\{\delta_0\}} \frac{\text{SPA}(F,n)}{\text{ECA}(F,n)} = 0.
        \end{align*}
    \end{theorem}

    \begin{proof}

    Since $F \in \mathcal{F}$, we have that $\lim_{u \to 0^+} R_F(u) = \lim_{u \to 0^+} \bar{R}_F(u) = \hat{\bar{R}}_F(u) < \infty$. We can then extend the definition of $R_F: (0,1] \to [0, \infty)$ to $R_F:[0,1] \to [0, \infty)$ where $R_F(0) :=\lim_{u \to 0^+} R_F(u)$, and can do the same for $\bar{R}_F$ and $\hat{\bar{F}}_F$.

    We denote $R_0$ to agree with $R$ everywhere except for that $R_0(0) = \lim_{q \to 0} R(q) < \infty$ since we restrict our attention to $F \in \mathcal{F}$.

    From continuity, that $\bar{R}_F(0) = \hat{\bar{R}}_F(0)$, and that $\bar{R}_F(1) = \hat{\bar{R}}_F(1)$, we will have that there exists countably many non-overlapping non-empty open intervals $\left\{(a_i, b_i)\right\}_{i =1}^{I}$ with $I \in \mathbb{N} \cup \{\infty\}$ such that
    \begin{align*}
        \bigcup_{i=1}^I (a_i, b_i) = \left\{u \in [0,1]: \hat{\bar{R}}_F(u) > \bar{R}_F(u)\right\},
    \end{align*}
    and, for any $i \in \mathbb{N} \cap [0, I]$,
    \begin{align*}
        \hat{\bar{R}}_0(a_i) =\bar{R}_0(a_i),
    \end{align*}
    and 
    \begin{align*}
        \hat{\bar{R}}_0(b_i) =\bar{R}_0(b_i).
    \end{align*}

    Thus, we have that
    \begin{align*}
        \frac{\text{OPT}(F, n)}{\text{LA}(F,n)}
        &=
        \frac{n(n-1)\int_{q=0}^1 (1-q)^{n-2} \hat{\bar{R}}_0(q)}{n(n-1)\int_{q=0}^1 (1-q)^{n-2} \bar{R}_0(q)}\\
        &=
        \frac{\int_{q=0}^1 (1-q)^{n-2} \hat{\bar{R}}_0(q)}{\int_{q=0}^1 (1-q)^{n-2} \bar{R}_0(q)}\\
        &\le
        \sup_{i=1}^I
        \frac{\int_{q=a_i}^{b_i} (1-q)^{n-2} \hat{\bar{R}}_0(q)}{\int_{q=a_i}^{b_i} (1-q)^{n-2} \bar{R}_0(q)}
    \end{align*}
    Next, we will bound 
    $\frac{\int_{q=a_i}^{b_i} (1-q)^{n-2} \hat{\bar{R}}_0(q)}{\int_{q=a_i}^{b_i} (1-q)^{n-2} \bar{R}_0(q)}$ for any $i \in \mathbb{N} \cap [0, I]$. 

    Consider an arbitrary $(a_i, b_i)$ for $i \in \mathbb{N} \cap [0, I]$. We recall that $\hat{\bar{R}}_0(a_i) =\bar{R}_0(a_i)$ and $\hat{\bar{R}}_0(b_i) =\bar{R}_0(b_i)$. For any $q \in (a_i, b_i)$, from monotonicity of $\bar{R}_0$, we have that
    \begin{align*}
        \bar{R}_0(q) \ge \bar{R}_0(a_i),
    \end{align*}
    and, from reverse monotonicity of $\frac{\bar{R}_0(x)}{x}$, we have that
    \begin{align*}
        \frac{\bar{R}_0(q)}{q} \ge \frac{\bar{R}_0(b_i)}{b_i}.
    \end{align*}

    Therefore, for any $q \in (a_i, b_i)$,
    \begin{align*}
        \bar{R}_0(q) \ge \max\oo{\bar{R}_0(a_i), \frac{\bar{R}_0(b_i)}{b_i}q},
    \end{align*}
    while
    \begin{align*}
        \hat{\bar{R}}_0(q) = \frac{q-a_i}{b_i-a_i} \cc{\bar{R}_0(b_i) - \bar{R}_0(a_i)} + \bar{R}_0(a_i).
    \end{align*}

    Thus, we can get the bound of
    \begin{align*}
        \frac{\text{OPT}(F, n)}{\text{LA}(F,n)}
        \le \sup_{
        \substack{0 \le A < B < \infty\\
        0 \le a < b \le 1}}
        \frac{\int_{q =a}^b (1-q)^{n-2} \cc{\frac{q-a}{b-a} \cc{B - A} + A } dq}
        {\int_{q =a}^b (1-q)^{n-2} \max\oo{A, \frac{B}{b}q} dq}.
    \end{align*}

    For any $0 \le A < B < \infty$ and $0 \le a < b \le 1$, we can define $c \in [0,1]$ such that $A = \frac{B}{b}c$. Thus, 
    \begin{align*}
        c = \frac{Ab}{B} \in [a,b).
    \end{align*}
    We further define $t:=\frac{c-a}{b-a} \in [0,1)$, $\delta := \frac{a}{b-a} \in [0, \infty)$, $\lambda := \frac{b-a}{1-a} \in (0,1]$, so
    \begin{align*}
        \frac{\int_{q =a}^b (1-q)^{n-2} \cc{\frac{q-a}{b-a} \cc{B - A} + A } dq}
        {\int_{q =a}^b (1-q)^{n-2} \max\oo{A, \frac{B}{b}q} dq}
        &=
        \frac{\frac{Bb}{b-a} \int_{x=0}^1 \cc{\delta + t+(1-t) x} (1-a)^{n-2}(1-\lambda x)^{n-2} dx}{\frac{Bb}{b-a} \int_{x=0}^1 \cc{\delta + \max(t,x)} (1-a)^{n-2}(1-\lambda x)^{n-2} dx}\\
        &=
        \frac{\int_{x=0}^1 \cc{\delta + t+(1-t) x} (1-\lambda x)^{n-2} dx}{\int_{x=0}^1 \cc{\delta + \max(t,x)} (1-\lambda x)^{n-2} dx}\\
        &\le
        \frac{\int_{x=0}^1 \cc{t+(1-t) x} (1-\lambda x)^{n-2} dx}{\int_{x=0}^1 \max(t,x) (1-\lambda x)^{n-2} dx}.
    \end{align*}

    Therefore, we can get the bound of
    \begin{align*}
        \frac{\text{OPT}(F, n)}{\text{LA}(F,n)}
        &\le
        \sup_{\substack{t\in [0,1)\\
        \lambda \in (0,1]}}
        \frac{\int_{x=0}^1 \cc{t+(1-t) x} (1-\lambda x)^{n-2} dx}{\int_{x=0}^1 \max(t,x) (1-\lambda x)^{n-2} dx}\\
        &=
        \sup_{\substack{t\in [0,1)\\
        \lambda \in (0,1]}}
        \frac{\int_{x=0}^1 \cc{t+(1-t) x} \cc{(1-\lambda) +\lambda(1-x)}^{n-2} dx}{\int_{x=0}^1 \max(t,x) \cc{(1-\lambda) +\lambda(1-x)}^{n-2} dx}\\
        &=
        \sup_{\substack{t\in [0,1)\\
        \lambda \in (0,1]}}
        \frac{\int_{x=0}^1 \cc{t+(1-t) x} \cc{\sum_{k=0}^{n-2} {n-2 \choose k}(1-\lambda)^{n-2-k}\lambda^k(1-x)^k} dx}{\int_{x=0}^1 \max(t,x) \cc{\sum_{k=0}^{n-2} {n-2 \choose k}(1-\lambda)^{n-2-k}\lambda^k(1-x)^k} dx}\\
        &=
        \sup_{\substack{t\in [0,1)\\
        \lambda \in (0,1]}}
        \max_{k=0}^{n-2}
        \frac{\int_{x=0}^1 \cc{t+(1-t) x} {n-2 \choose k}(1-\lambda)^{n-2-k}\lambda^k(1-x)^k dx}{\int_{x=0}^1 \max(t,x) {n-2 \choose k}(1-\lambda)^{n-2-k}\lambda^k(1-x)^k dx}\\
        &=
        \sup_{\substack{t\in [0,1)\\
        \lambda \in (0,1]}}
        \max_{k=0}^{n-2}
        \frac{\int_{x=0}^1 \cc{t+(1-t) x} (1-x)^k dx}{\int_{x=0}^1 \max(t,x) (1-x)^k dx}.
    \end{align*}
    Computing the integral yields
    \begin{align*}
        \frac{\text{OPT}(F, n)}{\text{LA}(F,n)}
        &\le
        \sup_{\substack{t\in [0,1)\\
        \lambda \in (0,1]}}
        \max_{k=0}^{n-2}
        \frac{\frac{1+(k+1)t}{(k+1)(k+2)}}{\frac{(k+2)t+(1-t)^{k+2}}{(k+1)(k+2)}}\\
        &=
        \max_{k=2}^{n}
        \sup_{t\in [0,1)}
        \frac{1+(k-1)t}{kt+(1-t)^k}\\
        &=
        \max_{k=2}^{n}
        \alpha_k\\
        &=
        \alpha_n.
    \end{align*}
    Thus, we have shown that $\inf_{F \in \mathcal{F} -\{\delta_0\}} \frac{\text{ECA}(F,n)}{\text{OPT}(F,n)} \le \frac{1}{\alpha_n}$. Next, for $n \ge 2$, we will show that this is tight by providing a construction of a sequence of distribution $F_i$ such that
    \begin{align*}
        F_i = \oo{1-\frac{t^*_n}{i}}\delta_1 + \frac{t^*_n}{i}\delta_{i}.
    \end{align*}
    This will lead to $R_{F_i}$ converging pointwise to $R(u) := t^*_n\mathbf{1}_{u=0}+u$, $\bar{R}_{F_i}$ converging pointwise to $\bar{R}(u) := \max(t^*_n, u)$, and $\hat{\bar{R}}_{F_i}$ converging pointwise to $\hat{\bar{R}}(u) := t^*_n + (1-t^*_n)u$.

    The comparison between SPA and ECG can be done by considering a sequence of distribution $F_i$ such that $F_i = \oo{1-\frac{1}{\sqrt{i}}}\delta_1 + \frac{1}{\sqrt{i}}\delta_{i}$.

    \end{proof}

    Thus, we have that the competitive ratio
    \begin{align*}
        \frac{\text{ECA}(F;n)}{\text{OPT}(F;n)}
    \end{align*}
    is uniformly bounded across $F \in \mathcal{F}-\{\delta_0\}$ and $n \in \mathbb{N}$ by
    \begin{align*}
        1-e^{-c} \approx 0.682156
    \end{align*}
    where $c$ is the unique positive solution of $e^c=c+2$.

\bibliographystyle{alpha}
\bibliography{egbib}

@inproceedings{ronen2001approximating,
  title={On approximating optimal auctions},
  author={Ronen, Amir},
  booktitle={Proceedings of the 3rd ACM conference on Electronic Commerce},
  pages={11--17},
  year={2001}
}

@article{sornwanee2026very,
  title={Very Credible Auction},
  author={Sornwanee, Thanawat},
  journal={arXiv preprint arXiv:2609.14263},
  year={2026}
}

@article{myerson1981optimal,
  title={Optimal auction design},
  author={Myerson, Roger B},
  journal={Mathematics of operations research},
  volume={6},
  number={1},
  pages={58--73},
  year={1981},
  publisher={INFORMS}
}

\end{document}